# Safeguarding the safeguards

How best to promote AI alignment in the public interest

December 2023

AUTHORS
Oliver Guest – Research Analyst (IAPS)
Michael Aird — Affiliate (IAPS)
Seán Ó hÉigeartaigh – Programme Director of AI:FAR (CSER)

# Abstract

AI alignment work is important from both a commercial and a safety lens. With this paper, we aim to help actors who support alignment efforts to make these efforts as effective as possible, and to avoid potential adverse effects. We begin by suggesting that institutions that are trying to act in the public interest (such as governments) should aim to support specifically alignment work that reduces accident or misuse risks. We then describe four problems which might cause alignment efforts to be counterproductive, increasing large-scale AI risks. We suggest mitigations for each problem. Finally, we make a broader recommendation that institutions trying to act in the public interest should think systematically about how to make their alignment efforts as effective, and as likely to be beneficial, as possible.

# Executive summary

AI alignment work is important from both a commercial and a safety lens. With this paper, we aim to help actors who support alignment efforts to make these efforts as effective as possible and to avoid potential adverse effects. *(To section)*

- "Alignment" refers to the extent to which an AI system, or AI systems in general, act in a way that the developer intended.[1]
- We use "alignment efforts" to mean technical work to promote alignment, as well as attempts to support this work, e.g., by providing funding.
- We are particularly aiming to inform "public interest institutions" that might support alignment efforts. This is our term for organizations that are trying to act in the public interest, such as governments, international organizations, and philanthropic foundations.

We begin by suggesting that public-interest institutions should aim to support specifically alignment work *that reduces accident or misuse risks*. *(To section)*

- This goal will closely overlap with, but sometimes differ from, supporting alignment work in general. Alignment work not only reduces risks, but also makes AI systems straightforwardly more useful.
- More useful AI systems are often beneficial to society, but other actors (e.g., corporate AI developers) are already strongly incentivized to increase the usefulness of AI systems. As a result, alignment efforts that primarily increase usefulness are less likely to be neglected; public-interest institutions might add little marginal value if they focus on it. These efforts might even just replace work that would otherwise have been done by corporate developers.[2]

We then describe four problems which might cause alignment efforts to be counterproductive, increasing large-scale AI risks:

1. **Relative progress problem:** To minimize large-scale AI risks, society should aim for a high ratio of progress in AI safeguards, relative to progress on AI capabilities. Some alignment efforts may upset this ratio by causing more progress on capabilities than on safeguards. *(To section)*
2. **False sense of security problem:** Some alignment efforts might contribute to a false sense of security that the accident and misuse risks from a given AI system, or AI systems in general, are low. This might

---

[1] We discuss alternative definitions in the appendix.
[2] There could also be some risks that result from increasing the usefulness of AI systems. We discuss these in the context of the relative progress problem and the bad principal problem.

mean that various actors do not take steps that they would otherwise take to reduce these risks. *(To section)*
3. **Dangerous valley problem:** Because of externalities, increasing the alignment of a given AI system might push it into a dangerous "valley" where the system is sufficiently safe that the developer is incentivized to deploy it, but insufficiently safe to make this deployment beneficial from society's perspective. *(To section)*
4. **Bad principal problem:** If the actor with which an AI system is aligned ("the principal") wants to misuse the system, then increasing alignment would be harmful via helping this actor to perpetrate misuse. *(To section)*

In our discussion of each problem, we recommend ways to mitigate the problem:

| Name | Selected recommendations |
| --- | --- |
| **Relative progress problem** | *Public interest institutions should focus on alignment for large-scale risk reduction rather than alignment in general.* Processes such as consulting external experts and formalized red-teaming of given alignment efforts might be helpful for achieving this.<br><br>*Reduce risks from "dual-use" alignment findings,* such as by considering sharing some findings only with groups that are working on reducing risks from AI systems. National AI safety institutes might be well-placed to facilitate this information exchange. |
| **False sense of security problem** | *Better understand the alignment of a given AI system.* For example, use external auditors to assess alignment and understand that important alignment issues might be difficult to detect.<br><br>*Better understand the state of alignment efforts in general.* For example, promote norms around communicating whether a given alignment effort will remain helpful as AI systems become more capable.<br><br>*Preserve the ability to correct ill-advised deployments*, i.e., to change how or whether an AI system is deployed, should evidence of significant alignment issues emerge. |
| **Dangerous valley problem** | *Governance mechanisms to better align the level of risk that is accepted from a given AI deployment with society's interests.* For example, licensing regimes where regulators could block specific AI deployments if these deployments pose more risk than society is willing to accept. |

|  | *Limiting the number of actors that can unilaterally deploy AI systems in high-risk ways.* For example, placing strong information security around model weights so that reckless actors cannot steal and deploy unsafe AI systems. |
|---|---|
| **Bad principal problem** | *Prevent malicious actors from having powerful AI systems that are aligned to them.* We discuss two key steps:<br>● Governments should prevent malicious actors from being able to develop their own very capable AI systems, e.g., by blocking their access to large compute clusters.<br>● Prevent malicious actors from removing safeguards on existing powerful AI systems, e.g., with limits on the open-sourcing of the most powerful models.<br><br>*Deter or mitigate misuse.* For example, via criminal sanctions or via "hardening" society against misuse such as by batching cyber vulnerabilities. |

We then make a broader recommendation that public-interest institutions should think systematically about how to make their alignment efforts as effective, and as likely to be beneficial, as possible. *[(To section)](#)*

- For example, these institutions should assess the following questions when evaluating a given alignment effort: To what extent would the alignment effort reduce accident and misuse risks? To what extent would the alignment effort increase these risks, e.g., via one of the four problems? Can any increases be mitigated, e.g., with the specific recommendations made in this paper?
- We expect that it would be helpful for public-interest institutions to work with external experts to answer these questions. We give some suggestions of how to do this.

Alignment efforts are extremely valuable and should be a priority for many actors. Although we are concerned about the problems that we describe, we expect that these problems could be sufficiently mitigated to make public-interest institutions' efforts to support alignment highly beneficial.

# Table of Contents

# Introduction

Large-scale risks from AI have received significant attention in 2023. Hundreds of AI scientists and other notable figures have signed a statement that "Mitigating the risk of extinction from AI should be a global priority alongside other societal-scale risks such as pandemics and nuclear war" (CAIS, 2023). According to a YouGov poll, 70% of Americans agree with the same statement (AIPI, 2023). In the Bletchley Declaration, many world governments, including those of China, the EU, the UK, and the US stated that "there is potential for serious, even catastrophic, harm, either deliberate or unintentional, stemming from the most significant capabilities of [frontier] AI models" (UK Government, 2023).[3]

This widespread concern has led to increased interest from a wide range of actors in technical and governance measures to reduce large-scale risks from AI accidents and misuse. This paper focuses on a key measure: alignment of AI systems. We use "alignment" to mean "the extent to which an AI system, or AI systems in general, act in a way that the developer intended."[4] Note that being aligned with the developer often overlaps with being aligned with the user. For example, with some exceptions, such as around illegal or harmful content, a company such as OpenAI generally wants its AI systems to be as helpful as possible to end users.

We use "alignment efforts" to connote three overlapping kinds of work:
1. **Technical work that directly increases the alignment of AI systems**, e.g., researching or applying alignment techniques.
2. **Technical work that would not itself increase alignment but that might prevent misaligned AI systems from being developed or deployed**. Examples of this work include mechanistic interpretability, model safety evaluations, and work on "model organisms of misalignment."[5]

---

[3] Other notable examples that indicate concern among various groups include Bilge et al. (2023), Bengio et al. (2023a), Bengio et al. (2023b), Future of Life Institute (2023), and Guterres (2023).
[4] "Alignment" is defined by researchers in many ways; we discuss alternative definitions in the appendix. We focus on alignment to the developer rather than the end user because the former seems in practice to be more common, at least for the most advanced AI systems. For example, companies that provide chatbots to the public generally impose some limits on what the AI system will do for end users, such as not producing illegal or hateful content (see, e.g., *Usage policies*, 2023).
[5] Mechanistic interpretability is the study of taking a trained neural network and analyzing the weights to reverse engineer the algorithms learned by the model. This could be helpful for detecting whether an AI system is being deceptive (Räuker et al., 2023, p. 11). Model safety evaluations attempt to assess whether a given AI system has or

3. **Attempts to support the other two lines of work**, e.g., by providing them with funding or institutional backing.

Alignment efforts should be a priority for many actors. In this paper, we aim to help actors who support alignment efforts to make these efforts as effective as possible, and to avoid potential adverse effects. We are primarily aiming to inform institutions that might spend significant resources on alignment efforts and that are aiming to act in the public interest. We refer to these collectively as "public-interest institutions." Examples include:

- **Government agencies**. See, for example, specific funding programs from the [NSF](#) in the US, and [ARIA](#) in the UK.
- **International organizations**. See, for example, proposals for a "CERN for AI safety" (Anthis, 2023; Ho et al., 2023, pp. 13–14; Marcus, 2023).
- **Philanthropic foundations**. See, for example, the second and third "hard problems in AI" that [Schmidt Futures](#) hopes to solve. See also the efforts by [ten allied philanthropies](#) to ensure that AI advances the public interest, including by improving interpretability and transparency.

We begin by noting that alignment work does not just have the effects of reducing AI risks, but also of making AI systems more useful. Insofar as these effects come apart, we argue that public-interest institutions should specifically be aiming to promote the former. We give examples of how to do so. Efforts that focus on large-scale risk reduction are more likely to be an effective use of resources and may avoid some downside risks.

We then describe four possible problems which might cause public-interest alignment efforts to be counterproductive, i.e., to increase large-scale risks. For each problem, we suggest specific mitigations:[6]

1. **Relative progress problem:** To minimize large-scale AI risks, society should aim for a high ratio of progress in AI safeguards, relative to progress on AI capabilities. Some alignment efforts may upset this ratio by causing more progress on capabilities than on safeguards.

---

will have dangerous capabilities and/or a propensity to apply these capabilities for harm. Concerning evaluation results might indicate that an AI system should not be further developed or deployed, at least without further alignment work (Shevlane et al., 2023). Model organisms are "in vitro demonstrations of the kinds of failures that might pose existential threats" from AI systems. They could be helpful for improving our understanding of potential failures, as well as for informing policy discussions around misalignment (Hubinger et al., 2023).

[6] We order the problems roughly according to how much we expect them to cause adverse effects, though we have not made a rigorous attempt to rank them, and we do not justify our rankings here.

2. **False sense of security problem:** Some alignment efforts might contribute to a false sense of security that the accident and misuse risks from a given AI system, or AI systems in general, are low. This might mean that various actors do not take steps that they would otherwise take to reduce these risks.[7]
3. **Dangerous valley problem:** Because of externalities, increasing the alignment of a given AI system might push it into a dangerous "valley" where the system is sufficiently safe that the developer is incentivized to deploy it, but insufficiently safe to make this deployment beneficial from society's perspective.
4. **Bad principal problem:** If the actor with which an AI system is aligned ("the principal") wants to misuse the system, then increasing alignment would be harmful via helping this actor to perpetrate misuse.

As well as making recommendations specific to each problem, we finish with a more general discussion. We suggest ways in which public-interest institutions can work systematically to make their alignment efforts as effective, and as likely to be beneficial, as possible.

Although we are concerned about the problems that we describe, we are confident that these can be sufficiently mitigated or avoided to make alignment efforts overall very beneficial. We hope that various public interest institutions support alignment efforts while bearing these recommendations in mind.

---

[7] One such measure could be a licensing regime for frontier AI development that would aim to reduce the likelihood of very powerful but misaligned AI systems being developed (Anderljung et al., 2023).

# Alignment efforts that prioritize risk reduction

Alignment work can both reduce the large-scale risks from AI systems and make AI systems more useful, such as by making AI systems more likely to produce an output that is helpful to the user. Indeed, the same specific techniques often do both. Insofar as these effects come apart, we claim in this section that public-interest institutions should prioritize alignment efforts that have the highest ratio of reducing large-scale risks to increasing usefulness.[8] This is primarily because these efforts are more likely to be neglected by other actors. Additionally, these efforts may have fewer downside risks.

Our claim does not necessarily imply that public-interest institutions should not support work that has significant effects on usefulness. If these institutions follow our strategy, then the more that they spend on alignment efforts, the more they will pursue opportunities to support alignment efforts with a particularly high ratio of risk reduction to increasing usefulness. If these institutions are spending significant resources on alignment efforts, this might mean that their best option for what to support next would be an alignment effort that has a relatively low ratio.

In this section, we discuss these considerations in more detail and give examples of alignment efforts that might have a particularly high ratio of reducing large-scale risks to increasing usefulness.

## Effects of alignment work

Alignment work could reduce at least two classes of large-scale AI risks:[9]
- **Accident risks:** An AI system might act in a way that no one intends, and thus pose hazards. For example, goal-directed AI systems might implement goals that no person would want, or act in an adversarial way in order to pursue these goals, such as by attempting to deceive humans (Bengio et al., 2023a, pp. 2–3; Chan et al., 2023, pp. 13–14, UK Department for Science, Innovation & Technology, 2023, pp. 8, 16, 27–28). If an AI system is more aligned to its developer, then it is less likely to have goals that no one would want or to act in an adversarial way as a result.[10]

---

[8] By "public-interest alignment efforts," we mean alignment efforts enacted by institutions that are trying to act in the public interest, e.g., government funding programs. This contrasts with alignment efforts that might be pursued by self-interested actors such as profit-maximizing corporate AI developers – even if the alignment efforts of corporate AI developers would also have positive externalities for society more broadly.

[9] For more complete taxonomies of AI risks, see Hendrycks et al. (2023), Maham & Küspert (2023), or Zwetsloot & Dafoe (2019). We leave out of scope here structural risk. This consists of risks from how AI shapes the broader (competitive) environments in which it is used (Zwetsloot & Dafoe 2019). That said, we expect that many of the problems and mitigations discussed in this paper could also be applied to structural risk.

- **Misuse risks:** Many AI systems are used by end users that are different from their developer. The end user might want to use an AI system for something that constitutes misuse by the lights of the developer. For example, a chatbot user might want to use the bot to learn how to build a bioweapon, despite the bot's developer not wanting the bot to help with this kind of request.[11] In such cases, the more that an AI system is aligned (to the developer), the more likely it is to reject misuse requests.[12] Note that if an AI system is aligned to its developer, that system could still be misused *by the developer*. We call this the "bad principal problem" and discuss it below.

Alignment does not purely have the effect of reducing large-scale AI risks; if an AI system is more likely to do what its developer intends, then that system is straightforwardly more useful for the developer (Khlaaf 2023, pp. 6–8). For example, consider a chatbot that a company develops for its internal use. If this chatbot answers questions in a way that is more helpful to the company, then the chatbot is more useful to the developer because it is more aligned. Chatbots are often used primarily by people other than the developer; for example, most ChatGPT users are not OpenAI employees. A similar logic applies in these cases;

---

[10] Note that, at least in some taxonomies (e.g., Hendrycks & Mazeika, 2022, p. 4), there can be accident risks (e.g., robustness failures) that result from technical problems other than inadequate alignment. We leave these other accident risks out of scope but expect that many of the points made in this paper would also apply to these risks.

[11] See, for example, concerns expressed by Anthropic and OpenAI that their products might be used for bioterror (Amodei, 2023, pp. 3–4; OpenAI, 2023, pp. 12–13).

[12] One could imagine cases where a given alignment effort contributes more to making an AI system helpful to end users (regardless of whether they have good intentions), rather than to making the AI system harmless by the lights of the developer. Such alignment efforts might *increase* misuse risk. For example, GPT-4 was first aligned with a round of RLHF to make it as helpful as possible to users (creating "GPT-4-early"), before then undergoing a round of RLHF to make it more harmless. A red-teamer who experimented with GPT-4-early describes the model as follows (Labenz, 2023): "It did its absolute best to satisfy the user's request – no matter how deranged or heinous your request! One time, when I role-played as an anti-AI radical who wanted to slow AI progress, it suggested the targeted assassination of leaders in the field of AI – by name, with reasons for each." If OpenAI had released GPT-4-early, then this would have been more aligned in some sense than the original GPT-4, but in a way that increased misuse risk. Note that no one is suggesting that OpenAI ever intended to release GPT-4-early as the finished product.

a chatbot is more useful to the developer if that chatbot is (generally) more useful to the end user.[13] This is because end users will be more willing to pay developers for access to AI systems that the users find valuable.[14]

Indeed, the same alignment techniques often both reduce risks and increase usefulness. For example, RLHF can be used both to make AI systems better at summarizing text (Stiennon et al., 2020a, 2020b), as well as more likely to decline harmful requests (Bai et al., 2022).[15] Similarly, Irving et al. (2018) propose "AI safety via debate" as a way for AI systems "to learn complex human goals and preferences." The authors both introduce the technique in terms of safety and describe how the technique can improve accuracy on classifying images.[16]

## The case for prioritizing risk reduction in public-interest alignment efforts

Having more useful AI systems is desirable, at least if the increased usefulness does not increase AI risks. That said, we claim that public-interest alignment efforts should focus on lines of work that have the highest ratio of reducing-large scale risks to increasing usefulness. There are two reasons for prioritizing risk reduction in this way:

1. **Efforts that aim particularly at large-scale risks might be more neglected by other actors**.[17] This is because corporate AI developers already have particularly strong incentives to make their systems more useful.[18] We discuss the neglectedness point immediately below.

---

[13] We write that developers only "generally" want their AI systems to be helpful to users. This is to account for cases where the end user wants to use the chatbot in a way that the developer views as harmful, and so where the developer might not want the system to be as helpful as possible to this user. For example, many chatbots refuse to use racist language, even if the end user wants them to.

[14] Developers might also find it helpful in other ways to have end users find their products useful. For example, this might increase the prestige or degree of market capture of the developer. Higher usefulness might also cause people to use the product more and thereby produce more training data that the developer can use to train subsequent AI systems.

[15] An additional complication is that RLHF might be most effective for alignment as a building block for more sophisticated techniques, rather than on its own (Christiano, 2023).

[16] Another example is Iterated Distillation and Amplification, which aims to align AI systems to complex values, and which is similar to the technique used to make AlphaGo more capable at playing Go (Christiano, 2017; Cotra, 2018).

[17] As noted above, there are close links between alignment to increase usefulness and alignment to reduce risks. Our claim is about how public-interest institutions should prioritize insofar as these types of alignment come apart.

[18] By "corporate AI developers" we mean AI developers that are not primarily motivated by the public interest. We focus on actors, such as corporations, that are motivated by profit. That said, there could be other types of actors. For example, military AI

2. **There are some ways in which increasing the usefulness of AI systems might be harmful**; we discuss these in our sections on the relative progress problem and bad principal problem. That said, we expect that it will often be beneficial to align AI systems in a way that makes them more useful, even when taking downside risks into account.

There are several reasons to think that alignment efforts focused on large-scale risk reduction are likely to be disproportionately neglected by corporate AI developers, relative to alignment efforts focused on usefulness.
- **Competitive dynamics** might make it rational for individual AI developers to underspend on reducing large-scale risks in order to use resources most effectively to get a competitive advantage over other developers (Askell et al., 2019, pp. 8–10; Critch & Krueger, 2020, p. 30; Hendrycks et al., 2023, pp. 16-18).
- **Visibility of capabilities relative to riskiness.** It might be easier for AI developers and others to observe and/or measure whether an AI system has become more useful or capable, relative to whether the large-scale risks from that system have decreased (Cotra, 2022; Duan, 2022). As a result, riskiness might be insufficiently salient to developers or other actors. Similarly, the lack of clear targets like benchmarks might make it disproportionately difficult to make progress on reducing risks, e.g., by making it less clear when one is making progress (Hendrycks & Mazeika, 2022, p. 6, 35; Sinha & Croak, 2023).[19]
- **Externalities.** Corporate AI developers might externalize (in the economic sense) more of the risks of AI development than the benefits. For example, AI developers might capture much of the value created by their AI systems, without bearing an equivalent proportion of the impact from any catastrophic risks associated with their AI development (O'Keefe et al., 2020, p. 24). If so, then they would be disproportionately incentivized to work on usefulness. That said, we have not tried to assess the extent to which AI developers internalize benefits or risks from their systems.

---

development might increase the relative capabilities of one country's military, relative to other countries. This would presumably benefit that country (and its allies) rather than countries in general.

[19] Progress on usefulness often has short, direct feedback loops (e.g., reported experience of users, performance on well-defined benchmarks). Progress on reducing large-scale risk might be harder to measure, e.g., because it depends on more complex processes, rarer circumstances, and interactions with real-world environments that are unlikely to be available during the development stage (Duan, 2022; Shevlane et al., 2023, pp. 12–13). That said, there are some benchmarks that are particularly relevant to large-scale risks, such as "MACHIAVELLI," which evaluates "agents' tendencies to be power-seeking, cause disutility, and commit ethical violations" (Pan et al., 2023).

If risk reduction alignment efforts are indeed disproportionately neglected by corporate AI developers, then public-interest alignment efforts are likely to be a particularly effective use of resources if they focus on risk reduction. One reason for this is that there is likely to be more remaining "low-hanging fruit" in this area, meaning that marginal additional resources go further. An additional reason is that alignment efforts that focus on risk reduction might be less likely to simply replace alignment efforts that would otherwise have been done by corporate developers anyway.

## How to prioritize risk reduction?

Some specific alignment efforts might be particularly helpful for reducing large-scale risks, relative to increasing usefulness. Hendrycks & Woodside (2022) describe several relevant lines of effort, including:

- **Power-averseness:** avoiding AI systems that would attempt to gain more power than is necessary.[20] This might reduce the likelihood or severity of AI accidents.
- **Honest AI:** reducing the likelihood of AI systems being deceptive, making it easier to tell whether these systems are acting or would act in a harmful way. Note that some work on this topic would be very helpful for usefulness, e.g., by making language models more likely to answer everyday questions accurately.
- **Anomaly detection:** detecting when an AI system is behaving in a novel way, potentially indicating that new hazards have emerged (relating to accidents, misuse, or something else).

Another alignment effort that might particularly reduce large-scale risks relative to increasing usefulness is:

- **Model organisms of misalignment** (Hubinger et al., 2023): developing "in vitro" demonstrations of the most concerning alignment failures to improve our understanding of them. This might have various risk-reducing effects, such as helping society to better target subsequent alignment efforts.

Despite having given these examples, we do not recommend that public-interest institutions *just* focus on these lines of effort. There are two reasons for this.

1. Experts might disagree about which alignment efforts have the highest ratio of reducing large-scale risks to increasing usefulness, and the answer might depend on the specific details.[21] As a result, we suggest that public-institutions think systematically about which alignment efforts are most beneficial to support. We suggest how to do this in the final section of the paper.
2. Opportunities to support alignment efforts with a particularly high ratio might already be taken, depending on how many resources are going towards public-interest alignment. This means that the next best option for alignment efforts might be efforts that only have a somewhat favorable ratio between reducing large-scale risks and increasing usefulness.

---

[20] Carlsmith (2022, pp. 18-22) gives a detailed description of what power-seeking AI might look like. Carlsmith cites examples such as AIs attempting to ensure self-preservation and to acquire various kinds of resources.

Having argued that public-interest institutions should be particularly focused on alignment that reduces large-scale risks, we now map four problems which could cause these alignment efforts to *increase* these risks. We make recommendations for mitigating each of them.

---

[21] Indeed, reviewers sometimes disagreed with the examples that we gave here.

# 1. Relative progress problem

Some alignment efforts might accelerate the development of subsequent more capable AI systems. We define an AI system as more capable if that system is able to perform better at some set of tasks, or if it achieves the same average performance on a wider set of tasks.

In many cases, it will be beneficial to have more capable AI systems. That said, more capable AI systems are more likely to have potentially dangerous capabilities (Anderljung et al., 2023, p. 9); these could pose new or increased large-scale risks, e.g., from accidents or misuse. As a result, progress on potentially dangerous capabilities would ideally be outpaced by progress on technical and governance safeguards to reduce these risks. Increased alignment is one such safeguard; as discussed earlier, it reduces some accident and misuse risks. Other safeguards might include governance measures to reduce the likelihood of poorly aligned systems being developed, or steps to "harden" society against accidents or misuse.[22]

We briefly discuss four mechanisms through which alignment efforts might accelerate the development of subsequent more capable AI systems. If these mechanisms have a large enough effect, then they might cause progress on potentially dangerous capabilities to outpace progress on safeguards. In this case, the relevant alignment efforts might have been counterproductive via doing more to accelerate risks from dangerous capabilities than to bolster safeguards to reduce these risks.[23]

---

[22] One example of a governance measure that might reduce the likelihood of poorly aligned systems being developed is a licensing regime for frontier AI development (Anderljung et al., 2023). An example of "hardening" could be increasing cybersecurity to defend against AI-enabled cyberattacks (Anderljung & Scharre, 2023).

[23] For a related discussion of the point that safety interventions can also improve capabilities, and that safety research thus needs to improve safety *relative to general capabilities*, see Hendrycks et al. (2023, pp. 29–30). Some readers might object that alignment work focused on reducing catastrophic risks may be easier and more effective to carry out in future, when AI systems will be more capable than today. Reasons for thinking this include that researchers will have a more informed sense of what the particularly risky AI systems look like, and that society will be able to use advanced AI systems to assist with alignment efforts. If one accepts this view, then it seems less valuable to increase the *current* ratio of safety to capabilities progress. Note, however, that even under this view it will be valuable at some point in the future to maximize the ratio of safety to capabilities, meaning that the considerations in this section might still be important at that point.

# More useful existing AI systems accelerating capabilities progress

Alignment efforts might increase the usefulness of existing AI systems, even if those alignment efforts are focused specifically on reducing large-scale risks. That is for at least two reasons.[24] First, as discussed in the previous section, the same alignment techniques often both reduce risks and increase usefulness. Second, if the risks from a given AI system are lower, then the developer will feel comfortable deploying or using it in a wider range of cases.[25]

Increasing the usefulness of existing AI systems might accelerate capabilities progress in at least two ways:[26]

- **Usefulness increasing investment in capabilities:** More useful systems are likely to be more profitable to their developer, increasing the resources that the developer can invest in training the next generation of systems. For example, because RLHF made GPT-4 more useful (Casper et al., 2023, p. 2), users might be more willing to pay OpenAI for access to GPT-4. The correspondingly higher profitability increases the resources that OpenAI can invest in subsequent training runs.[27] Relatedly, the existence of more useful AI systems might lead other actors to choose to allocate additional resources to AI development, or to allocate resources to this area for the first time.[28]
- **Usefulness contributing to AI development.** Existing AI systems can be helpful for the research and development of subsequent more capable AI systems: AI systems are already used to some extent to help improve the compute, algorithms, and data that are necessary for producing AI systems (UK Department for Science, Innovation & Technology, 2023, p. 15; Woodside, n.d.).[29] More useful existing AI systems would accelerate this effect.

---

[24] An additional possible reason is that alignment efforts may reduce regulatory burdens that would otherwise make it difficult to deploy AI systems. For example, multiple jurisdictions are considering rules that would require AI decision-making in certain contexts to be explainable (Nannini et al., 2023). Advances in mechanistic interpretability could be helpful for meeting these requirements, lowering the regulatory barriers to deploying a given AI system.
[25] See the "Dangerous valley problem" section for a more detailed discussion of this.
[26] We thank Jakob Graabak for his input on these points.
[27] Higher profitability might leave the developer with more resources to invest in subsequent training runs and/or make it easier for the developer to raise capital for the next training run.
[28] For example, Alphabet reportedly declared a "code red" after the release of ChatGPT, dramatically increasing its focus on AI product development (Grant, 2023).
[29] For detail on the claim that AI development depends upon compute, algorithms, and data, see Buchanan (2020).

## Capabilities insights from alignment efforts

Insights from alignment efforts could incidentally also generate insights that are helpful for building more capable systems. For example, mechanistic interpretability research might be important for alignment by making it easier to detect AI deception, but could also produce insights into how to design more capable model architectures (Räuker et al., 2023, pp. 10-11). Indeed, Poli et al. (2023) write that their work on increasing context length "would not have been possible" without inspiration from mechanistic interpretability work. That said, it might be surprising if incidental capabilities insights are a major effect of alignment efforts; there are already many actors deliberately trying to build more capable AI systems, meaning that incidental effects from alignment efforts on capabilities might be small on the margin.

## Resources for alignment efforts being deliberately used to increase capabilities

Researchers who want to advance capabilities might find ways to access resources that public-interest institutions intend to spend on alignment efforts. For example, researchers seeking funding from public-interest actors might be able to frame their work as about risk reduction and/or alignment, despite their research primarily being about increasing capabilities. This phenomenon may be difficult to avoid for at least two reasons. First, reasonable people can disagree about whether a given piece of work primarily benefits alignment relative to capabilities; a given piece of work will often have effects on both (Hendrycks & Mazeika, 2022, pp. 7–9). Second, these types of work often require similar specialized resources, e.g., compute and machine learning expertise.[30]

The likelihood of alignment resources being used primarily for capabilities advancements seems especially high if large amounts of funding are given out through programs with broad scopes and/or unclear definitions of specific safety or alignment issues.[31]

---

[30] For example, Jan Leike, who co-leads the "superalignment" team at OpenAI, highlighted this similarity in an interview, particularly from around 00:13:00 and 02:25:00 (Wiblin and Harris, 2023).

[31] We have some concerns that relatively undetailed proposals for a "Manhattan project for AI safety" or a "CERN for AI safety" (e.g., Anthis 2023; Ho et al., 2023, pp. 13–14; Marcus 2023) might be particularly likely to lead to this outcome. That said, well-executed versions of such efforts might be extremely beneficial, and it is understandable that proposals would only start with a high-level sketch of what the end result should look like.

## Alignment efforts increasing interest in building more capable AI systems

Discussions of alignment, particularly if focusing on large-scale misuse or accident risk, might increase people's beliefs that future AI systems could be extremely capable; arguments that AI could be very risky often involve a claim that AI could be very powerful. This could backfire if it results in actors becoming more focused on developing more powerful AI systems, perhaps out of a fear of being left behind while others develop powerful systems, without being sufficiently cautious of the risks. Indeed, Anthropic, (Google) DeepMind, and OpenAI, three organizations leading the advancement of AI capabilities, were each founded, to varying degrees, because of concerns about large-scale AI risks (Matthews, 2023; Metz, 2023; Perrigo, 2023a).[32] That said, the potential power of AI currently seems to receive a lot of public attention in any case.[33] As a result, the effect on the margin of this mechanism may be small.

## Recommendations

To minimize large-scale AI risks, society should aim for a high ratio of progress in AI safeguards, relative to progress on potentially dangerous AI capabilities. However, as this section has highlighted, some alignment efforts may upset this ratio by causing more progress on capabilities than on safeguards. Several approaches might be helpful for addressing this.

First, insofar as these effects come apart, public-interest institutions could aim to support alignment that specifically reduces risks rather than alignment efforts in general; much of the capabilities acceleration from alignment efforts that we have described results from alignment that increases usefulness. Given the close links between these lines of work, it may be difficult for institutions to determine which possible alignment efforts best meet this criterion. We suggest two possible solutions:
- **It might be helpful for institutions to consult with outside experts to gain additional opinions on the main effects of a given alignment effort**.[34] The authors may be able to recommend experts that would be knowledgeable in particular cases.
- **Projects that are seeking or using resources for alignment efforts could be expected to explain in detail why that project reduces risks in**

---

[32] We do not attempt to assess here how to what extent each of these organizations contributes to capabilities relative to alignment and/or safety.
[33] See many of the examples cited in the first paragraph of this paper.
[34] One list of such experts is compiled by the AI Safety Communications Centre, though we do not claim that this list is comprehensive.

**particular.** This explanation should be subjected to critical scrutiny, such as with formalized "red-teaming," or with norms among alignment experts that encourage such scrutiny. Hendrycks & Mazeika (2022, pp. 7–9, 21–34) make a similar recommendation and provide several detailed examples of what this could look like.

Second, if a given alignment finding is "dual-use," i.e., could be helpful both for reducing large-scale risks, as well as for accelerating progress towards potentially dangerous capabilities, decision-makers could consider disseminating this finding in a way that disproportionately contributes to risk reduction. For example, they could consider only sharing dual-use findings with people who are working directly on implementing safeguards around powerful AI systems. That said, this practice would go against norms of openness and reproducibility in science, and would presumably hamper the ability to scrutinize and build on the finding. As such, this will often not be an appropriate strategy. Organizations that have expressed some intention to facilitate this kind of information sharing include the UK's AI Safety Institute and the Frontier Model Forum (GOV.UK, 2023; Frontier Model Forum, 2023).

Third, it may be possible to reduce the phenomenon where talking about alignment increases others' interest in building AI systems with potentially dangerous capabilities. For example:
- Various mechanisms could make actors feel less need to recklessly develop high-risk AI systems. These include commitments from AI developers to share the benefits from powerful AI systems (e.g., O'Keefe et al., 2020), or not to use AI systems for military purposes.[35]
- It may be helpful, in some contexts, to avoid focusing AI risk discussions on the military applications of AI, such as for cyberattacks or detecting nuclear submarines (Maas et al., 2022). National security is often framed in a zero-sum way,[36] and so is particularly likely to promote discussions about who develops risky AI systems, rather than whether these systems are made safe.[37] Additionally, AI development in national security

---

[35] Precedents of commitments around military AI include pledges by governments not to use AI in specific military contexts (e.g., Rautenbach, 2023), as well as pledges by AI developers not to develop lethal autonomous weapons (Future of Life Institute, 2018).
[36] See, for example, discussion of the "security dilemma" (Rittberger, 2004, pp. 3–4).
[37] Additionally, researchers could highlight true information that might make military or defense decision-makers reluctant to use AI in national security contexts. For example, AI systems are often non-robust and non-reliable in ways that are especially important in high-impact and sensitive domains such as the military (Horowitz & Scharre 2021, pp. 7–9).

contexts might be particularly dangerous, even if only thinking about risk from accidents.[38]

Finally, many interventions by governments, AI developers, or other actors could reduce large-scale risks from AI systems, even assuming the existence of AI systems that are significantly more capable than today's. See, for example, the proposals in Anderljung et al. (2023), Anderljung & Hazell (2023), and Schuett et al. (2023).

---

[38] For example, it may be harder to achieve safety-promoting governance measures in the military context, and deployments of military AI might be particularly dangerous because they would involve giving the AI system access to dangerous tools (Trager et al., 2023, p. 14). That said, we expect that a lot of AI accident risk is not specific to national security applications and instead results from characteristics that other AI systems might have, such as generality and situational awareness (Ngo et al., 2023).

# 2. False sense of security problem

Alignment efforts might cause a false sense of security among "deployment decision-makers," or society more broadly, that the risk of large-scale accidents or misuse from a given AI system, or from AI in general, is low. By "deployment decision-makers" we mean people who make decisions about whether to deploy AI systems. Examples of deployment decision-makers might include senior figures in organizations that develop powerful AI systems, or AI regulators. There are at least three ways in which the false sense of security problem could occur.[39]

First, alignment work efforts cause deployment decision-makers to wrongly believe that a given AI system is more aligned than it is. Alignment efforts might cause AI systems to superficially appear aligned, without in fact having this property. This concern applies to both accident and misuse risks.[40]

- **Accidents**: People can make mistakes about whether a given alignment effort has succeeded. For example, Amodei et al. (2017) describe a case where a robot which was supposed to grasp items instead positioned its manipulator in between the camera and the object so that it appeared to be grasping the item, fooling human evaluators. A particularly concerning version of this threat model is "deceptive alignment," where an agentic and misaligned model would deliberately act as if aligned during training in order to be deployed so that it could pursue its misaligned goals (Carlsmith, 2023).[41]
- **Misuse**: Deployment decision-makers might incorrectly believe that sufficient safeguards to prevent misuse have been added, only for malicious users to find unexpected ways to evade these, such as jailbreaks.[42]

---

[39] A possible fourth way in which alignment efforts might contribute to a false sense of security is if these efforts only address some risks associated with a given AI system, and deployment decision-makers do not realize that other risks remain. As a hypothetical example, if one aligns a language model such that it is less likely to give advice on using viruses for bioterror, this might create a false sense of security if the model would still give advice on using other pathogens for bioterror, and deployment decision-makers do not realize this.

[40] For a more detailed discussion of why deployment decision-makers might initially misjudge the safety of an AI system, as well as what they should do in this case, see O'Brien et al. (2023).

[41] Discussions around deceptive misalignment are currently hypothetical; existing AI systems do not seem to have the capabilities to enact such sophisticated deception strategies.

[42] For a detailed discussion of jailbreaks, see Wei et al. (2023). One possible example of deployment decision-makers underestimating misuse risk is Microsoft's 2016 "Tay"

Incorrect beliefs about how aligned AI systems are could be harmful; they could cause insufficiently aligned systems to be deployed, when they would not be if relevant decision-makers were better informed.[43]

Second, there could be a false sense of security from alignment work that reduces large-scale risks from *existing* AI systems to very low levels, without the techniques scaling well to *future* AI systems; future systems might be more agentic or capable and so require more advanced safeguards (Chan et al., 2023; Critch & Krueger, 2020, p. 38). This phenomenon could cause harm if it causes society to believe that large-scale AI risks are generally low, when in fact the risks are just *temporarily* low. This belief could prevent society from implementing measures to reduce the risks from future AI systems, such as policy measures to make it less likely that very capable but misaligned AI systems are developed or deployed (Anderljung et al., 2023). As an example, it appears that Microsoft could have used existing alignment techniques, in particular, RLHF to make the Bing/Sydney system less blatantly misaligned (Miles & Goodside, 2023). However, it may have been good for the world that Microsoft failed to do so. These existing techniques would not necessarily have scaled to more capable AI systems (Casper et al., 2023, pp. 6–8). As such, using them on Bing/Sydney might have contributed to a false sense of security about future AI systems. In contrast, the case led to more widespread awareness of alignment concerns (see e.g., Perrigo, 2023b), without causing much harm.

Third, the likelihood of a false-sense of security might be particularly high if some AI developers are engaging in "safety-washing" (de Freitas Netto et al., 2020, p. 6; Hendrycks et al., 2023, p. 30).[44] Analogous to "greenwashing," this is a possible phenomenon where AI developers might do work that is more motivated by *appearing* to be improving safety rather than in fact reducing risks from insufficient AI alignment.[45] Safety-washing might be beneficial (at least in the short term) to AI developers, perhaps by improving their reputation, or

---

chatbot. Online trolls managed to get Tay to output many abusive Tweets, causing Microsoft to decide to take Tay down (Schwartz, 2019).

[43] Note that accidental or misuse catastrophes might result from AI systems, even before a lab decides to deploy that system. As such, the considerations here might apply to decisions about whether to start or continue a training run, not just whether to deploy the resulting model (Anderljung et al., 2023, pp. 20-21).

[44] There are additional reasons why safety-washing might be harmful. For example, safety-washing would make it harder for society to incentivize good alignment work by making it less clear what work is valuable. We thank Lizka Vaintrob for contributing to our understanding of safety-washing.

[45] There may be a particular risk of safety-washing because, as described above, there is often an overlap between alignment techniques that make AI systems more useful and alignment techniques that make AI systems safer.

helping them to avoid safety-motivated regulation or boycotts.[46] However, safety-washing might be harmful by making it even harder for deployment decision-makers or society to have an accurate sense of the risks from a given AI system or AI in general. This might prevent society from implementing necessary governance measures to reduce large-scale risks.

## Recommendations

Three classes of recommendations might help mitigate the false sense of security problem. First, actors including AI developers, external auditors, and perhaps governments should make it easier to have an accurate sense of the alignment of a given AI system. Second, these actors should make it easier to have an accurate sense of the state of alignment efforts in general. Third, AI developers should maintain the ability to correct the way in which AI systems are deployed, e.g., if they incorrectly judged the alignment of that system. We discuss these recommendations immediately below.

### Better understand the alignment of a given AI system

Various steps could help deployment decision-makers to understand the level of alignment of a given AI system, making them less likely to overestimate how aligned it is:

- External auditors, not just the AI developer, should assess the alignment of high-risk AI systems. External actors could provide an additional perspective on whether a system is sufficiently aligned. They might also differ from AI developers in that they are less strongly incentivized to find that a system can be safely deployed (Mökander et al., 2023, pp. 8–9; Raji et al., 2022, pp. 3-4). Various labs have already committed to having their models audited for safety (Ee & O'Brien, 2023), and auditing could be mandated via regulation (Anderljung et al., 2023, pp. 20–21, 26).
- Because very capable AI systems might be dangerous even if they superficially appear to be aligned, deployment decisions should be made based on evidence that is likely to reveal any serious alignment issues, including hard-to-detect issues such as "deceptive" misalignment.

### Better understand the state of alignment efforts in general

There are also several steps that could reduce the likelihood of society having an incorrect sense of AI risks in general:

---

[46] It may also be easier or cheaper for developers to do work that appears to be helpful for alignment rather than work that fundamentally is helpful for alignment.

- If a given alignment effort reduces the risks from existing AI systems but will not necessarily do so for more capable AI systems, then communications about the alignment effort should make this clear.
- To improve discussions about alignment, it might be helpful for researchers to develop more precise breakdowns and definitions of alignment.[47] Institutions that are distributing resources for alignment should expect researchers to be precise about the kinds of alignment that they are researching. For example, researchers could be expected to describe which hazards they have addressed, and whether they expect the work to also address these hazards in more capable AI systems.[48]
- To make decision-makers less vulnerable to safety-washing, outside experts, perhaps commissioned by government agencies, could assess claims about a given piece of alignment research and publicize their assessments.[49]

Preserve the ability to correct ill-advised deployments

Despite the steps above, deployment decision-makers might overestimate the alignment of a given system. This could lead to them deploying an AI system that they subsequently realize poses unacceptably high accident or misuse risk. Developers should preserve the ability to make "deployment corrections," i.e., to change how or whether a model is deployed. For example, they could make models available via API, allowing them to subsequently correct how the model is deployed if they realize that a system is more dangerous than initially thought (O'Brien et al., 2023, especially pp. 10–14).

---

[47] One example is that OpenAI has now introduced the term "superalignment" to refer specifically to efforts to align very capable AI systems as opposed to existing models (Leike & Sutskever 2023).
[48] Hendrycks & Mazeika (2022, pp. 7–9, 21–34) make a similar recommendation and provide several detailed examples of what this could look like.
[49] So that there is an incentive towards high-quality alignment work, experts should be willing to praise developers for alignment efforts that are genuinely helpful, even if those experts also express worries about the behavior of that developer in general.

# 3. Dangerous valley problem

Perhaps counterintuitively, increasing the alignment of a given AI system might increase the likelihood of large-scale accidents or misuse from that system via increasing the probability of that system being deployed. Because of externalities, increasing the alignment of a given AI system might push it into a dangerous "valley" of safety levels where the system is sufficiently aligned to make deployment beneficial to the developer, without being safe enough to make this deployment beneficial to society in general. We call this the dangerous valley problem.[50]

Consider an AI system that might be very beneficial to the developer if deployed, e.g., because it would be very profitable, but that also carries some risk from large-scale accidents or misuse:[51]

- If alignment is very low, then the likelihood of a post-deployment accidental or misuse catastrophe is very high. In this case, the developer would not deploy the system, even if entirely self-interested; the risk to the developer from the catastrophe outweighs the potential benefit to the developer from deployment. In this scenario, society is safe on one side of the valley.[52]
- If alignment is very high, then society is mostly safe on the other side of the valley; the developer is likely to deploy the system, but this is unlikely to lead to large-scale accidents or misuse by end users. Note, however, that in this case the *developer* could still misuse the AI system, e.g., if the developer has malicious intentions.[53]
- At an intermediate level of alignment, the risk from deploying the system might be low enough that it is in the developer's self-interest to deploy the system, but not low enough that it is in society's collective interests for the system to be deployed.[54] An alignment effort that gets an AI system to this level (and not beyond this level) would have pushed society into the dangerous valley.

---

[50] We would like to thank David Krueger for improving our understanding of this problem.
[51] In this scenario, we are assuming that the developer has a reasonably accurate sense of the risk level from deploying a given AI system. Note that, due to uncertainty and subjective judgment over the likelihood and pathways to large-scale accidents, and potential uncertainty over the robustness of alignment techniques, the dangerous valley might be entered *unwittingly* by a developer.
[52] If an AI system has very low alignment, then developers might avoid deploying it not just because that system poses high risks, but also because the system is unlikely to be useful even if deployed.
[53] We discuss misuse by AI developers in our discussion of the bad principal problem.
[54] The phenomenon that an AI risk reduction measure might make an AI developer more likely to accept the remaining risk is an example of "risk compensation." See discussion of this in the AI context in Stafford et al. (2022, pp. 3–4) and more generally in Reynolds (2014, pp. 177–178).

The dangerous valley problem does not just apply to AI developers, but rather to any actor that influences the decision about whether to deploy a model with some accident or misuse risk. These could include regulators and people who might be able to put social pressure on developers, such as the broader machine learning community and the public. Additionally, there might not only be a dangerous valley at the point where a model is trained and can then either be deployed or not deployed. Sufficiently capable models might be dangerous from an accident and misuse perspective even during training or limited deployment (Anderljung et al., 2023, pp. 20-21; ARC Evals 2023, p. 10). This means that there is also a dangerous valley problem at the point where actors decide whether to start or continue training an AI system.

## Recommendations

Governance mechanisms might somewhat better align the AI developer's risk tolerance with society's, or cause the AI developer not to deploy when that is against society's interests (even if the developer wants to). For example:

- With a licensing regime, AI developers could be required to seek a license to develop and/or deploy risky AI models (Anderljung et al., 2023, pp. 20–21). If the body responsible for granting the license were a government, its incentives might be more aligned with society's interests than private developers' incentives are.[55]
- A "Windfall Clause" (O'Keefe et al., 2020), where AI developers redistribute profits above some very high threshold, might help reduce the extent to which AI developers internalize AI's benefits more than its risks.

Additionally, various decision-makers can limit, or at least maintain the ability to limit, the proliferation of cutting-edge AI development and deployment capabilities. This would reduce the number of actors that could unilaterally push society into the valley. For example:[56]

- Governments could restrict access to very large clusters of computational resources ("compute"), reducing the number of actors who are able to produce models powerful enough to push humanity into the valley. For example, governments could impose licensing regimes on large compute clusters or place export controls on AI-relevant hardware (Trager et al., 2023, especially pp. 26–28).
- Strong information security around the weights of the most high-risk AI systems would make these systems harder to steal, reducing the number of actors that could choose to deploy these systems (Anderljung et al., 2023, p. 21). AI developers should increase this information security and could be supported in doing so by governments (Ee & O'Brien, 2023).

---

[55] For example, governments might benefit less from a given AI system being deployed than that system's developer would. Governments might also be aiming to be more representative of society than private AI developers are, particularly when the AI developer has obligations to shareholders.

[56] We note that policies to reduce proliferation will often have trade-offs that have to be carefully considered. For example, restricting access to large compute clusters might raise concerns about concentration of power.

# 4. Bad principal problem

If the actor with which an AI system is aligned ("the principal") wants to misuse the system, then increasing alignment would be harmful via helping this actor to perpetrate misuse.[57] For example, totalitarian governments that have AI systems aligned to them might be able to use these systems to entrench their power or otherwise cause harm, such as by using these systems for mass surveillance (Brundage et al., 2018, p. 47; Dafoe, 2018, p. 7).[58] In particular, alignment might help malicious principals by reducing the risks to the principals of using these systems,[59] as well as by making the AI system more helpful to the principal.

Note that, for many AI systems, the principal is not the same as the actor that might misuse the system. For example, discussions about the misuse of GPT-4 often focus on how end users might misuse it (see, e.g., Gaulkin, 2023). In contrast, to the extent that GPT-4 is aligned, it is aligned with OpenAI, i.e., the developer.[60] In cases where the principal is different to the user that might commit misuse, there is no bad principal problem; increasing the alignment of the system would reduce the likelihood of misuse because the principal would not want the system to be misused.

## Recommendations

Following Anderljung & Hazell (2023), we group possible recommendations to address misuse into which part of the "misuse chain" they target. We focus only on recommendations that would help mitigate the bad principal problem, where the misuse is perpetrated by the principal, rather than AI misuse in general.

---

[57] We mean "principal" here in the same sense as in discussions of the principal-agent problem. The "principal" is an individual or entity who delegates work to another to perform services on the principal's behalf (Oxford Reference, 2009).
[58] We chose here examples that would widely be recognized as misuse and that could also be seen as malicious. However, one might reasonably also define misuse to include cases where the principal is trying to act ethically and legally but is mistaken about how to do so.
[59] For example, if an AI system is insufficiently aligned and thus prone to catastrophic accidents, then actors (including malicious principals) will be less willing to deploy or use that AI system.
[60] Alignment with the developer and alignment with the end user often overlap because the developer generally wants AI systems to be helpful to the end user. These types of alignment come apart in cases where the developer does not want the AI system to do things that the end user wants it to do. For example, OpenAI tries to prevent ChatGPT from producing some harmful content, even if end users want this content (*Usage Policies*, 2023).

Before misuse

Society could act before misuse occurs by ensuring that malicious actors do not have access to powerful AI systems that are aligned to them. One step here is to prevent malicious actors from being able to develop their own powerful AI systems. For example, governments could establish licensing regimes for developing powerful AI systems and deny licenses to actors that are likely to be malicious. Similarly, governments could control access to very large compute clusters; these are currently necessary for developing powerful AI systems. That said, producing the most powerful AI systems requires significant technical expertise and costs tens of millions of dollars, at least currently (Anderljung et al., 2023, p. 13).[61] As a result, there are currently relatively few actors that would be able to produce powerful systems even without licensing and compute controls.

We expect that a more important step is preventing malicious actors from having full access to powerful AI systems developed by others; if one has access to the weights of the model, it is relatively easy and cheap to make a copy of the system that is aligned to oneself rather than the initial developer. For example, it only requires hundreds of dollars and consumer-grade computers to largely remove the safety features added by developers to large language models, *if one has the model weights* (Lermen et al., 2023).[62] This implies that it might be desirable to restrict the number of people who have access to the weights of very powerful AI systems. Measures to do this include improving the information security practices of AI developers (Anderljung et al., 2023, p. 21), or considering limits on when model weights can be open-sourced (Seger et al., 2023, p. 12).[63]

These policies need not prevent the public from using powerful AI systems. Models can be released via structured access, where the developer does not share the weights but still allows others to use the model, such as via a web interface (Shevlane, 2022). This provides the AI developer greater scope for controlling how the AI system is used, including making it harder for end users to misuse the system. Various AI products, such as ChatGPT, take this approach. Note that, as "jailbreaks" demonstrate, AI developers often cannot fully control how their systems are used, even if these systems are provided via structured access (Wei et al., 2023).

---

[61] Without intervention, the cost of producing models of a given capability level is likely to decrease over time due to improvements in the hardware and algorithmic inputs (Erdil & Besiroglu 2022; Hobbhahn et al., 2023).

[62] Fine-tuning models via API, i.e., without oneself having access to the weights, can also undo alignment in this way. That said, there are various techniques that developers can use to reduce the risks from this and that developers can enforce if they control access to the weights (Qi et al., 2023, especially pp. 9–12).

[63] We note that there are advantages as well as disadvantages to releasing model weights, and that there are many options between completely releasing and completely not releasing a powerful AI system to the public. We recommend Seger et al. (2023) for a detailed discussion of these considerations.

### During or after misuse

Various measures might further reduce misuse risk at other steps of the "misuse chain," i.e., while or after the misuse occurs. We only discuss these measures briefly because they are helpful for misuse in general rather than the bad principal problem specifically. We recommend Anderljung & Hazell (2023) for a more detailed discussion.

- **AI systems can sometimes be used to "harden" society against the misuse of other AI systems.** For example, it might be possible to use AI systems to patch cyber vulnerabilities before malicious actors misuse other AI systems to hack the vulnerable systems (Anderljung & Scharre, 2023).
- **Detecting and punishing misuse** (such as with the criminal justice system) might disincentivize actors from perpetrating misuse.

# Systematically choosing beneficial alignment efforts

In the sections above, we have made several specific recommendations for how public-interest institutions can support alignment efforts to reduce large-scale risks that are both effective and unlikely to be counterproductive.[64] That said, there is a lot of possible variation among possible alignment efforts. As such, our recommendations will not always apply, and the importance of the considerations that we have highlighted will vary from case to case.

As a result, our overarching recommendation is that public-interest institutions should think systematically about alignment efforts that they are considering supporting. In particular, these institutions should carefully assess the following questions.
- To what extent would a given alignment effort reduce accident and misuse risks?
- To what extent would the alignment effort increase these risks via one of the four problems, or via some different mechanism?
- Can any risk increases be mitigated, e.g., with the specific recommendations made in this paper?

Hendrycks & Mazeika (2022, pp. 7–9, 21–34) make a similar recommendation and provide detailed examples of what this could look like.

Given the difficulty of assessing the value of a given alignment effort, we expect that institutions would often benefit significantly from external expertise when making these assessments. For example:
- Institutions could reach out to relevant experts in AI safety and alignment topics for advice or red-teaming of their plans. [65]
- Institutions could publicly describe their approaches to supporting alignment efforts, so that a wide range of people can critique them. It might be helpful to have norms, e.g., among communities that focus on AI alignment or safety, that promote this kind of open critique.
- Projects that are seeking or using resources for alignment efforts could be expected to explain in detail why that project is particularly helpful for reducing large-scale risks. This explanation could be subjected to critical scrutiny, such as with formalized "red-teaming," or with norms among alignment experts that encourage such scrutiny.

---

[64] Some of our above recommendations also apply to actors other than institutions that are themselves trying to support alignment efforts, e.g., to regulators.
[65] One list of experts is compiled by the AI Safety Communications Centre, though we do not claim that this list is comprehensive. The authors may also be able to recommend additional experts that would be knowledgeable in specific cases.

Alignment efforts are extremely valuable and should be a priority for many actors. Although we are concerned about the problems that we describe, we expect that these problems could generally be sufficiently avoided or mitigated to make public-interest institutions' efforts to support alignment highly beneficial.

# Appendix

## Defining alignment

Researchers define "alignment" in various ways. Although we leave a full literature review out of scope, we highlight the following "clusters" of definitions:[66]

1.  **Minimal definitions.** The actions or intentions of AI systems align with the intentions or preferences of a specific actor, such as the developer or end user. Our definition belongs in this cluster. Examples:
    -   Christiano (2018): "AI A is aligned with an operator H [if and only if] A is trying to do what H wants it to do."[67]
    -   Leike et al. (2018, p. 1) describe the agent alignment problem: "How can we create agents that behave in accordance with the user's intentions?"
2.  **Maximal definitions.** The actions or intentions of AI systems align with producing morally good outcomes. Examples:
    -   Gabriel (2020, p. 12): "AI could be designed to align with [values]: the agent does what it morally ought to do, as defined by the individual or society."
    -   Russell (2020, p. 137): "The goal of AI value alignment is to ensure that powerful AI is properly aligned with human values."[68]
3.  **Hazard-focused definitions.** Alignment means avoiding some safety failures from AI systems. This might include catastrophic or even existential safety failures. Examples:
    -   Critch & Krueger (2020, p15): "We say that a prepotent AI system is misaligned if it is unsurvivable (to humanity), i.e., its deployment would bring about conditions under which the human species is unable to survive."[69]

---

[66] For somewhat more comprehensive reviews, we recommend Gabriel (2020) and Ngo (2020). We take the "minimal" and "maximal" terms from Gabriel (2020, p. 3), and were inspired by Hendrycks & Mazeika (2022, p. 4) for the "hazard-focused" term. For simplicity, we generally focus here on aligning a single AI system to a single actor ("single/single"). Critch & Krueger (2020) discuss in detail other scenarios: single/multi, multi/single, and multi/multi.

[67] Christiano clarifies that he means in the "de dicto" rather than "de re" sense; an AI is aligned if it does what it believes the operator wants it to do, even if that belief is incorrect.

[68] Note that Russell sometimes writes more in terms of hazard-focused definitions.

[69] The authors also use a minimal definition on the same page: "AI alignment refers to the problem of ensuring that an AI system will behave well in accordance with the values of another entity, such as a human, an institution, or humanity as a whole."

- Hendrycks & Mazeika (2022, p. 4): "Alignment research seeks to make AI systems less hazardous by focusing on hazards such as power-seeking tendencies, dishonesty, or hazardous goals."

In this paper, we say that an AI system is aligned to the extent that it acts in the way intended by the developer. We focus on alignment to the developer rather than the end user because the former seems in practice to be more common, at least for the most advanced AI systems. For example, companies that provide chatbots to the public generally impose some limits on what the AI system will do for end users, such as not producing illegal or hateful content (see, e.g., *Usage Policies*, 2023).[70] That said, note that being aligned with the developer often overlaps with being aligned with the user. For example, apart from in cases like illegal or harmful content, OpenAI presumably generally wants their models to be as helpful as possible to end users.[71]

---

[70] As another example, manufacturers of autonomous vehicles would presumably not allow the user to operate the vehicle in a malicious way, e.g., trying to hit pedestrians, even if the user wanted that.

[71] For more on the tradeoffs between helpfulness (to end users) and harmlessness, see Bai et al. (2022, especially pp. 4–5).

# Acknowledgements

We are grateful to the following people for discussion and input: Justus Baumann, Tamay Besiroglu, Ben Bucknall, Jakob Graabak, Erich Grunewald, John Halstead, José Hernández-Orallo, David Manheim, Sören Mindermann, Max Räuker, Tilman Räuker, Jaime Sevilla, Rohin Shah, Ben Stevenson, Zoe Williams, Caleb Withers, Lexin Zhou, and the participants of an IAPS seminar. Mistakes and opinions are our own. We are also grateful to Adam Papineau for copyediting.

# References

*AI Safety Fundamentals*. (n.d.). BlueDot Impact. Retrieved November 9, 2023, from https://perma.cc/54XR-GNXG

Amodei, D. (2023). *Written Testimony for a hearing on "Oversight of A.I.: Principles for Regulation" Before the Judiciary Committee Subcommittee on Privacy, Technology, and the Law*. https://perma.cc/74Z7-8CXK

Amodei, D., Christiano, P., & Ray, A. (2017, June 13). *Learning from human preferences*. OpenAI. https://perma.cc/VUT3-JR9Z

Anderljung, M., Barnhart, J., Leung, J., Korinek, A., O'Keefe, C., Whittlestone, J., Avin, S., Brundage, M., Bullock, J., Cass-Beggs, D., Chang, B., Collins, T., Fist, T., Hadfield, G., Hayes, A., Ho, L., Hooker, S., Horvitz, E., Kolt, N., … Wolf, K. (2023). *Frontier AI Regulation: Managing Emerging Risks to Public Safety* (arXiv:2307.03718). arXiv. http://arxiv.org/abs/2307.03718

Anderljung, M., & Hazell, J. (2023). *Protecting Society from AI Misuse: When are Restrictions on Capabilities Warranted?* (arXiv:2303.09377). arXiv. http://arxiv.org/abs/2303.09377

Anderljung, M., & Scharre, P. (2023, August 14). *How to Prevent an AI Catastrophe*. Foreign Affairs. https://perma.cc/C4NJ-N7GT

*Announcing Chris Meserole as Executive Director*. (2023, October 25). Frontier Model Forum. https://perma.cc/ZB7P-NPCK

Anthis, J. R. (2023, May 18). *Why we need a "Manhattan Project" for A.I. safety*. Salon. https://perma.cc/TM7N-3BY3

Askell, A., Brundage, M., & Hadfield, G. (2019). *The Role of Cooperation in Responsible AI Development* (arXiv:1907.04534). arXiv. http://arxiv.org/abs/1907.04534

Bai, Y., Jones, A., Ndousse, K., Askell, A., Chen, A., DasSarma, N., Drain, D., Fort, S., Ganguli, D., Henighan, T., Joseph, N., Kadavath, S., Kernion, J., Conerly, T., El-Showk, S., Elhage, N., Hatfield-Dodds, Z., Hernandez, D., Hume, T., … Kaplan, J. (2022). *Training a Helpful and Harmless Assistant with Reinforcement Learning from Human Feedback* (arXiv:2204.05862). arXiv. https://doi.org/10.48550/arXiv.2204.05862

Bengio, Y., Hinton, G., Yao, A., Song, D., Abbeel, P., Harari, Y. N., Zhang, Y.-Q., Xue, L., Shalev-Shwartz, S., Hadfield, G., Clune, J., Maharaj, T., Hutter, F., Baydin, A. G., McIlraith, S., Gao, Q., Acharya, A., Krueger, D., Dragan, A., … Mindermann, S. (2023a). *Managing AI Risks in an Era of Rapid Progress* (arXiv:2310.17688). arXiv. https://doi.org/10.48550/arXiv.2310.17688

Bengio, Y., Yao, A., Russell, S., & Zhang, Y.-Q. (2023b). *Prominent AI Scientists from China and the West Propose Joint Strategy to Mitigate Risks from AI*. Center for Human-Compatible Artificial Intelligence. https://perma.cc/S98B-9UBZ

Bilge, T., Lifland, E., Mann, J., Hickman, M., Campos, S., & Wasil, A. (2023). *Urging an International AI Treaty: An Open Letter*. AI Treaty. https://perma.cc/3J2C-K4DH

Brundage, M., Avin, S., Clark, J., Toner, H., Eckersley, P., Garfinkel, B., Dafoe, A., Scharre, P., Zeitzoff, T., Filar, B., Anderson, H., Roff, H., Allen, G. C., Steinhardt, J., Flynn, C., Ó hÉigeartaigh, S., Beard, S., Belfield, H., Farquhar, S., … Amodei, D. (2018). *The Malicious Use of Artificial Intelligence: Forecasting, Prevention, and Mitigation* (arXiv:1802.07228). arXiv. https://doi.org/10.48550/arXiv.1802.07228

Buchanan, B. (2020). *The AI Triad and What It Means for National Security Strategy*. Center for Security and Emerging Technology. https://doi.org/10.51593/20200021

*Capabilities and risks from frontier AI: A discussion paper on the need for further research into AI risk*. (2023). UK Department for Science, Innovation & Technology. https://perma.cc/49M9-KDWU

Carlsmith, J. (2022). *Is Power-Seeking AI an Existential Risk?* (arXiv:2206.13353). arXiv. http://arxiv.org/abs/2206.13353

Carlsmith, J. (2023). *Scheming AIs: Will AIs fake alignment during training in order to get power?* (arXiv:2311.08379). arXiv. http://arxiv.org/abs/2311.08379

Casper, S., Davies, X., Shi, C., Gilbert, T. K., Scheurer, J., Rando, J., Freedman, R., Korbak, T., Lindner, D., Freire, P., Wang, T., Marks, S., Segerie, C.-R., Carroll, M., Peng, A., Christoffersen, P., Damani, M., Slocum, S., Anwar, U., … Hadfield-Menell, D. (2023). *Open Problems and Fundamental Limitations of Reinforcement Learning from Human Feedback* (arXiv:2307.15217). arXiv. https://doi.org/10.48550/arXiv.2307.15217

Chan, A., Salganik, R., Markelius, A., Pang, C., Rajkumar, N., Krasheninnikov, D., Langosco, L., He, Z., Duan, Y., Carroll, M., Lin, M., Mayhew, A., Collins, K., Molamohammadi, M., Burden, J., Zhao, W., Rismani, S., Voudouris, K., Bhatt, U., … Maharaj, T. (2023). *Harms from Increasingly Agentic Algorithmic Systems*. 2023 ACM Conference on Fairness, Accountability, and Transparency, 651–666. https://doi.org/10.1145/3593013.3594033

Christiano, P. (2017, October 20). *AlphaGo Zero and capability amplification. AI Alignment*. https://perma.cc/7CEB-L3L8

Christiano, P. (2018, November 15). *Clarifying "AI Alignment."* AI Alignment Forum. https://perma.cc/6XK4-JP7J

Christiano, P. (2023, January 25). *Thoughts on the impact of RLHF research.* Alignment Forum. https://perma.cc/7JCE-J98K

Cotra, A. (2018, April 29). *Iterated Distillation and Amplification.* AI Alignment. https://perma.cc/5JMX-CDSM

Cotra, A. (2022, July 18). *Without specific countermeasures, the easiest path to transformative AI likely leads to AI takeover.* Alignment Forum. https://perma.cc/Y7JZ-Y8ZL

Critch, A., & Krueger, D. (2020). *AI Research Considerations for Human Existential Safety (ARCHES)* (arXiv:2006.04948). arXiv. https://doi.org/10.48550/arXiv.2006.04948

Dafoe, A. (2018). *AI Governance: A Research Agenda.* Centre for the Governance of AI. https://perma.cc/79WB-L2LS

Dalrymple, D. (2023). *Mathematics and modelling are the keys we need to safely unlock transformative AI.* Advanced Research + Invention Agency. https://perma.cc/VSZ7-BH7A

De Freitas Netto, S. V., Sobral, M. F. F., Ribeiro, A. R. B., & Soares, G. R. D. L. (2020). *Concepts and forms of greenwashing: A systematic review.* Environmental Sciences Europe, 32(1), 19. https://doi.org/10.1186/s12302-020-0300-3

Demski, A., & Garrabrant, S. (2020). *Embedded Agency* (arXiv:1902.09469). arXiv. https://doi.org/10.48550/arXiv.1902.09469

Duan, I. (2022, December 3). *Race to the Top: Rethink Benchmark-Making for Safe AI Development.* https://perma.cc/SZ5J-JS56

Ee, S., & O'Brien, J. (2023, August 4). *Putting New AI Lab Commitments in Context.* Centre for the Governance of AI. https://perma.cc/K7YH-W7LB

Erdil, E., & Besiroglu, T. (2022, December). *Revisiting Algorithmic Progress.* https://perma.cc/4AQL-5V49

Gabriel, I. (2020). *Artificial Intelligence, Values and Alignment.* Minds and Machines, 30(3), 411–437. https://doi.org/10.1007/s11023-020-09539-2

Gade, P., Lermen, S., Rogers-Smith, C., & Ladish, J. (2023). *BadLlama: Cheaply removing safety fine-tuning from Llama 2-Chat 13B* (arXiv:2311.00117). arXiv. https://doi.org/10.48550/arXiv.2311.00117

Gaulkin, T. (2023, March 30). *What happened when WMD experts tried to make the GPT-4 AI do bad things.* Bulletin of the Atomic Scientists. https://perma.cc/A77Z-HYFN

*GPT-4 System Card.* (2023). OpenAI. https://perma.cc/2TD8-LE5Z

Grant, N. (2023, January 20). *Google Calls In Help From Larry Page and Sergey Brin for A.I. Fight.* The New York Times. https://perma.cc/54FQ-7M6M

Guterres, A. (2023, July 18). *Remarks to the Security Council debate on artificial intelligence.* UN Press. https://perma.cc/6UV3-9J86

Hendrycks, D., & Mazeika, M. (2022). *X-Risk Analysis for AI Research* (arXiv:2206.05862). arXiv. https://doi.org/10.48550/arXiv.2206.05862

Hendrycks, D., Mazeika, M., & Woodside, T. (2023). *An Overview of Catastrophic AI Risks* (arXiv:2306.12001). arXiv. https://doi.org/10.48550/arXiv.2306.12001

Hendrycks, D., & Woodside, T. (2022, June 10). *Open Problems in AI X-Risk.* AI Alignment Forum. https://perma.cc/VFQ8-VSG9

Ho, L., Barnhart, J., Trager, R., Bengio, Y., Brundage, M., Carnegie, A., Chowdhury, R., Dafoe, A., Hadfield, G., Levi, M., & Snidal, D. (2023). *International Institutions for Advanced AI* (arXiv:2307.04699). arXiv. http://arxiv.org/abs/2307.04699

Hobbhahn, M., Heim, L., & Aydos, G. (2023, November 9). *Trends in Machine Learning Hardware.* Epoch. https://perma.cc/VF5R-6CQC

Horowitz, M., & Scharre, P. (2021, January 12). *AI and International Stability: Risks and Confidence-Building Measures.* CNAS. https://perma.cc/P6GW-YXSB

Hubinger, E., Schiefer, N., Denison, C., & Perez, E. (2023, August 8). *Model Organisms of Misalignment: The Case for a New Pillar of Alignment Research.* AI Alignment Forum. https://perma.cc/3HKL-WR29

*Introducing the AI Safety Institute.* (2023, November 2). GOV.UK. https://perma.cc/JX6E-W4VT

Irving, G., Christiano, P., & Amodei, D. (2018). *AI safety via debate* (arXiv:1805.00899). arXiv. https://doi.org/10.48550/arXiv.1805.00899

Khlaaf, H. (2023). *Toward Comprehensive Risk Assessments and Assurance of AI-Based Systems.* Trail of Bits. https://perma.cc/4G83-XV8C

Labenz, N. (2023, November 22). *Did I get Sam Altman fired from OpenAI? The Cognitive Revolution.* https://perma.cc/5ZDL-WUCC

Leike, J., Krueger, D., Everitt, T., Martic, M., Maini, V., & Legg, S. (2018). *Scalable agent alignment via reward modeling: A research direction* (arXiv:1811.07871). arXiv. https://doi.org/10.48550/arXiv.1811.07871

Leike, J., & Sutskever, I. (2023, July 5). *Introducing Superalignment.* OpenAI. https://perma.cc/U8FS-TLSJ

Lermen, S., Rogers-Smith, C., & Ladish, J. (2023). *LoRA Fine-tuning Efficiently Undoes Safety Training in Llama 2-Chat 70B* (arXiv:2310.20624). arXiv. https://doi.org/10.48550/arXiv.2310.20624

*Lethal Autonomous Weapons Pledge.* (2018, June 6). Future of Life Institute. https://perma.cc/SRP5-3VWZ

Maas, M. M., Matteucci, K., & Cooke, D. (2022). *Military Artificial Intelligence as Contributor to Global Catastrophic Risk.* SSRN Electronic Journal. https://doi.org/10.2139/ssrn.4115010

Maham, P., & Küspert, S. (2023). *Governing General Purpose AI — A Comprehensive Map of Unreliability, Misuse and Systemic Risks [Policy Brief].* Stiftung Neue Verantwortung. https://perma.cc/5U7L-MV62

Manyika, J. (2023, June). *Working List of Hard Problems in AI.* AI2050, a Program of Schmidt Futures. https://perma.cc/249C-X2LH

Marcus, G. (2023, October 26). *A CERN for AI and the Global Governance of AI. Marcus on AI.* https://perma.cc/EG3Q-W7Z7

MATS Program. (n.d.). *ML Alignment & Theory Scholars.* Retrieved November 9, 2023, from https://perma.cc/G8TZ-EGD2

Matthews, D. (2023, July 17). *The $1 billion gamble to ensure AI doesn't destroy humanity.* Vox. https://perma.cc/5V83-9N98

Metz, C. (2016, April 27). *Inside OpenAI, Elon Musk's Wild Plan to Set Artificial Intelligence Free.* Wired. https://perma.cc/42GK-TNEK

Miles, R., & Goodside, R. (2023, March 24). *Bing Chat Behaving Badly—Computerphile.* YouTube. https://perma.cc/3HYG-DFW3

Mökander, J., Schuett, J., Kirk, H. R., & Floridi, L. (2023). *Auditing large language models: A three-layered approach.* AI and Ethics. https://doi.org/10.1007/s43681-023-00289-2

Nannini, L., Balayn, A., & Smith, A. L. (2023). *Explainability in AI Policies: A Critical Review of Communications, Reports, Regulations, and Standards in the EU, US, and UK* (arXiv:2304.11218). arXiv. https://doi.org/10.48550/arXiv.2304.11218

Ngo, R. (2020, October 1). *AGI safety from first principles: Alignment.* Alignment Forum. https://perma.cc/44UN-EVVX

Ngo, R., Chan, L., & Mindermann, S. (2023). *The alignment problem from a deep learning perspective* (arXiv:2209.00626). arXiv. http://arxiv.org/abs/2209.00626

O'Brien, J., Ee, S., & Williams, Z. (2023). *Deployment corrections: An incident response framework for frontier AI models.* Institute for AI Policy and Strategy. https://perma.cc/V5QT-U9XY

O'Keefe, C., Cihon, P., Garfinkel, B., Flynn, C., Leung, J., & Dafoe, A. (2020). *The Windfall Clause: Distributing the Benefits of AI for the Common Good* (arXiv:1912.11595). arXiv. https://doi.org/10.48550/arXiv.1912.11595

Pan, A., Chan, J. S., Zou, A., Li, N., Basart, S., Woodside, T., Ng, J., Zhang, H., Emmons, S., & Hendrycks, D. (2023). *Do the Rewards Justify the Means? Measuring Trade-Offs Between Rewards and Ethical Behavior in the MACHIAVELLI Benchmark* (arXiv:2304.03279). arXiv. http://arxiv.org/abs/2304.03279

*Pause Giant AI Experiments: An Open Letter.* (2023, March 22). Future of Life Institute. https://perma.cc/4WSW-4YEM

Perrigo, B. (2023a, January 12). *DeepMind CEO Demis Hassabis Urges Caution on AI.* Time. https://perma.cc/5E2W-L4FK

Perrigo, B. (2023b, February 17). *Bing's AI Is Threatening Users. That's No Laughing Matter.* Time. https://perma.cc/2744-6XXL

*Philanthropies launch new initiative to ensure AI advances the public interest.* (2023, November 1). Ford Foundation. https://perma.cc/J8RC-Q7H2

Poli, M., Massaroli, S., Nguyen, E., Fu, D., Dao, T., Baccus, S. A., Bengio, Y., Ermon, S., & Ré, C. (2023, March 7). *Hyena Hierarchy: Towards Larger Convolutional Language Models.* https://perma.cc/RVU7-A6PY

*Poll Shows Overwhelming Concern About Risks From AI as New Institute Launches to Understand Public Opinion and Advocate for Responsible AI Policies.* (2023, August 9). AI Policy Institute. https://perma.cc/P3BZ-XUUX

*Principal–agent problem.* (2009). Oxford Reference. https://doi.org/10.1093/oi/authority.20110803100346712

Qi, X., Zeng, Y., Xie, T., Chen, P.-Y., Jia, R., Mittal, P., & Henderson, P. (2023). *Fine-tuning Aligned Language Models Compromises Safety, Even When Users Do Not Intend To!* (arXiv:2310.03693). arXiv. https://doi.org/10.48550/arXiv.2310.03693

Raji, I. D., Xu, P., Honigsberg, C., & Ho, D. E. (2022). *Outsider Oversight: Designing a Third Party Audit Ecosystem for AI Governance* (arXiv:2206.04737). arXiv. http://arxiv.org/abs/2206.04737

Räuker, T., Ho, A., Casper, S., & Hadfield-Menell, D. (2023). *Toward Transparent AI: A Survey on Interpreting the Inner Structures of Deep Neural Networks* (arXiv:2207.13243). arXiv. https://doi.org/10.48550/arXiv.2207.13243

Rautenbach, P. (2023, February 16). *Keeping humans in the loop is not enough to make AI safe for nuclear weapons.* Bulletin of the Atomic Scientists. https://perma.cc/2E9L-J7H7

*Response to Request for Comments on AI Accountability Policy.* (2023). ARC Evals. https://perma.cc/U4UH-SUYV

Reynolds, J. (2014). *A Critical Examination of the Climate Engineering Moral Hazard and Risk Compensation Concern.* SSRN Electronic Journal. https://doi.org/10.2139/ssrn.2492708

Rittberger, V. (2004). *Approaches to the study of foreign policy derived from international relations theories.* Institute for Political Science, University of Tübingen.

Russell, S. J. (2020). *Human compatible: AI and the problem of control.* Penguin Books.

*Safe Learning-Enabled Systems.* (2023, February 23). National Science Foundation. https://perma.cc/AVP4-ZUTC

Schuett, J., Dreksler, N., Anderljung, M., McCaffary, D., Heim, L., Bluemke, E., & Garfinkel, B. (2023). *Towards best practices in AGI safety and governance: A*
SAFEGUARDING THE SAFEGUARDS | 41

*survey of expert opinion* (arXiv:2305.07153). arXiv. http://arxiv.org/abs/2305.07153

Schwartz, O. (2019, November 25). *In 2016, Microsoft's Racist Chatbot Revealed the Dangers of Online Conversation.* IEEE Spectrum. https://perma.cc/RKJ2-PPBF

Seger, E., Dreksler, N., Moulange, R., Dardaman, E., Schuett, J., Wei, K., Winter, C., Arnold, M., Ó hÉigeartaigh, S., Korinek, A., Anderljung, M., Bucknall, B., Chan, A., Stafford, E., Koessler, L., Ovadya, A., Garfinkel, B., Bluemke, E., Aird, M., … Gupta, A. (2023). *Open-Sourcing Highly Capable Foundation Models: An evaluation of risks, benefits, and alternative methods for pursuing open-source objectives.* Centre for the Governance of AI. https://perma.cc/W6WZ-RNMZ

Shevlane, T. (2022). *Structured access: An emerging paradigm for safe AI deployment* (arXiv:2201.05159). arXiv. http://arxiv.org/abs/2201.05159

Shevlane, T., Farquhar, S., Garfinkel, B., Phuong, M., Whittlestone, J., Leung, J., Kokotajlo, D., Marchal, N., Anderljung, M., Kolt, N., Ho, L., Siddarth, D., Avin, S., Hawkins, W., Kim, B., Gabriel, I., Bolina, V., Clark, J., Bengio, Y., … Dafoe, A. (2023). *Model evaluation for extreme risks* (arXiv:2305.15324). arXiv. http://arxiv.org/abs/2305.15324

Sinha, A., & Croak, M. (2023, October 26). *Supporting benchmarks for AI safety with MLCommons.* Google Research. https://perma.cc/7KFQ-8VLJ

Stafford, E., Trager, R. F., & Dafoe, A. (2022). *Safety Not Guaranteed: International Strategic Dynamics of Risky Technology Races.* Centre for the Governance of AI. https://perma.cc/36WS-FTGJ

*Statement on AI Risk.* (2023, May 30). Center for AI Safety. https://perma.cc/KDU3-YHPG

Stiennon, N., Christiano, P., Ziegler, D., Lowe, R., Wu, J., Voss, C., & Ouyang, L. (2020, September 4). *Learning to summarize with human feedback.* OpenAI. https://perma.cc/V99M-WNCZ

Stiennon, N., Ouyang, L., Wu, J., Ziegler, D. M., Lowe, R., Voss, C., Radford, A., Amodei, D., & Christiano, P. (2022). *Learning to summarize from human feedback* (arXiv:2009.01325). arXiv. https://doi.org/10.48550/arXiv.2009.01325

*The Bletchley Declaration by Countries Attending the AI Safety Summit, 1-2 November 2023.* (2023, November 1). GOV.UK. https://perma.cc/H277-K554

Trager, R. F., Harack, B., Reuel, A., Carnegie, A., Heim, L., Ho, L., Kreps, S., Lall, R., Larter, O., Ó hÉigeartaigh, S., Staffell, S., & Villalobos, J. J. (2023). *International Governance of Civilian AI: A Jurisdictional Certification Approach.* Oxford Martin AI Governance Initiative in partnership with the Centre for the Governance of AI. https://perma.cc/VFB7-RZ3N

*Usage policies.* (2023, March 23). OpenAI. https://perma.cc/6LVS-VWAK

Wei, A., Haghtalab, N., & Steinhardt, J. (2023). *Jailbroken: How Does LLM Safety Training Fail?* (arXiv:2307.02483). arXiv. http://arxiv.org/abs/2307.02483

Wiblin, R., & Harris, K. (2023, August 7). *Jan Leike on OpenAI's massive push to make superintelligence safe in 4 years or less.* 80,000 Hours. https://perma.cc/G3TR-4LL6

Woodside, T. (n.d.). *Examples of AI Improving AI.* Retrieved September 29, 2023, from https://perma.cc/M8LS-FU35

Yang, X., Wang, X., Zhang, Q., Petzold, L., Wang, W. Y., Zhao, X., & Lin, D. (2023). *Shadow Alignment: The Ease of Subverting Safely-Aligned Language Models* (arXiv:2310.02949). arXiv. https://doi.org/10.48550/arXiv.2310.02949

Zwetsloot, R., & Dafoe, A. (2019, February 11). *Thinking About Risks From AI: Accidents, Misuse and Structure.* Lawfare. https://perma.cc/4PA6-9ASJ